\documentclass{raa_twocolumn}            % article version

\usepackage{graphicx,times}             %for PS/EPS graphics inclusion, new
\usepackage{natbib}
\usepackage{amssymb,amsmath}
\usepackage{xcolor}
\bibpunct{(}{)}{;}{a}{}{,}
\usepackage[colorlinks=true,linkcolor=blue,citecolor=blue,draft=false]{hyperref}

\usepackage{comment}
\usepackage{booktabs}
\usepackage{multirow}
\usepackage{tablefootnote}

\begin{document}

  \title{The data analysis pipeline for the Microchannel X-ray Telescope on board the SVOM mission
}
%   \subtitle{I. Place Your Subtitle Here}

   \volnopage{} %%preserved for Editor. DOn't remove!
   \setcounter{page}{1} %%starting page, preserved for Editor. DOn't remove!

   \author{P. Maggi
      \inst{1,*}\footnotetext{$*$Corresponding Author, these authors contributed equally to this work.}
   \and L. Michel
      \inst{1}
    \and D. G\"otz \inst{2}
    % Orde alphabetique ensuite
    \and S. Crepaldi\inst{4}
    \and A. Fort\inst{4}
    \and L. Kleiver\inst{1}
    \and A. Lorang\inst{1}
    \and K. Mercier\inst{4}
    \and M. Moita \inst{2}
    \and P. Guillout \inst{1}
    \and C. Motch \inst{1}
    \and F. Robinet \inst{3}
    \and A. Sauvageon \inst{2}
      %\footnotetext{ID: https:/orcid.org/0000-0002-0003-0004}
   }
%% Here is an example of three authors come from different institutes.
%% For single author or all the authors from an institute, use "\inst{}" only

   \institute{Observatoire Astronomique de Strasbourg, Université de Strasbourg, CNRS, 11 rue de l’Universite , 67000 Strasbourg, France; {\it pierre.maggi@astro.unistra.fr}
    \and
    AIM-CEA/DRF/Irfu/Departement d’Astrophysique, CNRS, Universite Paris-Saclay, Universite Paris Cite, Orme des Merisiers, Bat. 709, Gif-sur-Yvette, 91191, France
    \and
    Universit\'e Paris-Saclay, CNRS/IN2P3, IJCLab, 91405 Orsay, France    
    \and
    Centre National d’Etudes Spatiales, Centre spatial de Toulouse, 18 avenue Edouard Belin, 31401 Toulouse Cedex 9, France
    \\
\vs\no
   {\small Received 202x month day; accepted 202x month day}}

\abstract{
The Space-based multi-band astronomical Variable Objects Monitor (SVOM) mission was launched in June 2024. It is a joint Sino-French collaboration designed to detect, localize, and study gamma-ray bursts (GRBs) and other high-energy transients. Among its onboard instruments, the Microchannel X-ray Telescope (MXT) plays a central role by providing follow-up X-ray observations of GRB afterglows and other transient phenomena.
To ensure timely and accurate scientific exploitation of MXT observations, a dedicated ground processing pipeline has been developed. This pipeline automatically ingests raw event lists, performs calibration, background and time filtering, corrects instrumental effects, and produces science-ready data products such as images and light curves and spectra of detected sources.
In this paper, we describe the architecture and key components of the MXT data analysis pipeline, highlighting its modular design and integration within the broader SVOM ground segment. We also show results from real datasets, demonstrating the pipeline’s ability to meet the performance requirements of the mission.
\keywords{software: data analysis --- gamma-ray bursts --- X-rays: general -- X-rays: bursts}
}

   \authorrunning{P. Maggi et al.}            %author_head in even pages
   \titlerunning{The MXT data analysis pipeline}  % title_head in odd pages

   \maketitle
\section{Introduction}           %% first-level sections will be auto-capitalized
\label{sect:intro}

This paper presents the data analysis pipeline for the Microchannel X-ray Telescope (MXT) on board SVOM \citep{2016arXiv161006892W}. The primary purpose of the MXT \citep{2023ExA....55..487G} is to provide accurate and rapid localization of the gamma-ray burst (GRB) afterglow in the X-ray band (0.3--10~keV). The in-flight scientific performance of MXT are presented in G\"otz et al. (RAA, 2025, 25, this issue), while the camera and optics are described in details in Moita et al. and Feldman et al. (RAA, 2025, 25, this issue), respectively.

To meet the scientific objectives of the SVOM mission (Cordier et al., RAA, 2025, 25, this issue), MXT data are first analysed by the on-board software (Robinet et al., RAA, 2025, 25, this issue) to obtain and immediately disseminate the localization of potential afterglow of GRBs or other high-energy transient sources detected by the large field-of-view instruments ECLAIRS and GRM (Godet et al.; Sun et al., RAA, 2025, 25, this issue).

Downlinked data are subsequently analysed by the ground data analysis pipeline described in the present paper. The main purposes of the MXT pipeline is to provide SVOM users with \textit{i)} calibrated data, \textit{ii)} observation and source products (e.g. detected source lists, images, energy spectra, light curves), hereafter referred to as Science Data Products (SDPs), and \textit{iii)} high-level analysis scientific products for detected sources, such as flux, spectral parameters, variability measurements, etc. All calibrated data and scientific products are stored on the Scientific Database (SDB) of the French Science Centre (FSC, Louvin et al., RAA, 2025, 25, this issue) and can be used for further analysis with external tools. SVOM users may also run the pipeline offline, customising (some) parameters when scientifically relevant.

In this work we present this data analysis pipeline in detail. The origin and type of input data used are presented in Sect.~\ref{sect:datastreams}.  The general workflow architecture is described in Sect.~\ref{sect:workflow}. We then detail in Sect.~\ref{sect:stage_calibration} the various stages and tasks of the pipeline: data preparation, events calibration, source detection, and source product generation and analysis. Performances and caveats are discussed in Sect.~\ref{sect:discussion}.

\section{Data streams}
\label{sect:datastreams}

The MXT pipeline processes MXT data starting at the data level 1 (L1\,\footnote{SVOM data levels are derived from the IVOA standards \citep{2017ivoa.spec.0509L}, merging the IVOA Level 3 and Level 4 products into a single SVOM L3 level}). The main L1 input is an event list in FITS format, preprocessed at the FSC from decoded raw values (Level 0 data) downlinked to the ground. There are two data streams which trigger the MXT pipelines, so called X-band and X-VHF.

\subsection{X-band data}
\label{sect:datastreams_xband}
These data are downlinked via X-band antennae and contain the most complete information on individual events at camera level. 
Each event is time-tagged with the time of the frame in which it was detected, and assigned the position in raw detector coordinates (\texttt{RAWY,RAWZ}) of the center of the event. Individual X-ray events are reconstructed at FSC from individual hit pixel events, assigning them a \texttt{MULTIPLICITY} (number of pixel with collected charge above threshold in the event) and a \texttt{GRADE} coding the shape or pattern of the event: 0 for a single-pixel event, 1-4 for a double-pixel events, etc.
%(see \ref{}\todo{ref to MXT camera paper? or XRT} for the full grade description).
In the analysis only valid X-ray events are kept, i.e. single- to quadruple-pixel with events of grade 0 to 12. Finally, the X-band L1 event list also contain the raw spectral information in the form of the detector-measured quantity "pulse-height analyzer" (PHA)\,\footnote{So called for historical reason, and standard for X-ray analysis as per OGIP standard 92-002, \url{https://heasarc.gsfc.nasa.gov/docs/heasarc/caldb/docs/memos/cal_gen_92_002/cal_gen_92_002.html}}, i.e. the ADU level in each pixel of the 3x3 grid of the event grade, in the \texttt{PHAS} column. This ADU energy is calibrated and converted to physical energy in the calibration stage of the pipeline (Sect.~\ref{sect:stage_calibration_energy}).
Owing to the the limited number of available antenna contact, X-band data become available a relatively long time after the observation (several hours).

\subsection{X-VHF data}
\label{sect:datastreams_xvhf}
To permit quick analysis in certain programs \citep[Core Program, exceptional of multi-messenger Target of Opportunity, as described in][]{2016arXiv161006892W}, the MXT pipelines also processes so called X-VHF data, which are pre-processed at FSC from decoded VHF packets and transformed to FITS file having the same format as X-band data, allowing most pipeline modules (see Sect.~\ref{sect:workflow}) to be used for both data streams. The main input from X-VHF data is a photon list which iteratively concatenates the raw photon hits on the detector in the first 10 packets received through the VHF alert network (Cordier et al., RAA, 2025, 25, this issue). Reception of all 10 packets may be delayed depending on VHF antennae availability, and pipeline results can improve as the input photon list accumulate. Each such packet contains at most 35 photons, meaning that the exposure time represented by the full list of 350 photons may be very short (few seconds) in the case of a bright GRB afterglow.

VHF packets are severely bandwidth-limited and only the raw detector position is recorded, and no energy or grade/multiplicity information is available. Only crude timing information, the time at emission of a packet containing all 35 photons, is available.
Nevertheless, the X-VHF data are available in quasi-real time (after downlink and transfer delay plus pre-processing time) and allow a quick ground-based analysis by the MXT pipeline, primarily of source detection and localization, \textit{independently} of the on-board software analysis (Robinet et al., RAA, 2025, 25, this issue).

\begin{table*}[t]
    \centering
    \caption{List of existing pipeline instances and the program in which they operate. The analysis modules they contain are indicated by a cross "X". CP: Core Program; GP: General Program; ToO-MM: Target of Opportunity Multi-Messenger; ToO-EX: Target of Opportunity Exceptional.}
    \label{tab:pipelines}
    %\vspace{-1em}
    \begin{tabular}{l c c c c c c}
        \hline\noalign{\smallskip}
        \multirow{2}{*}{Instance} &  \multirow{2}{*}{Data stream} & \multirow{2}{*}{Program\textdagger}
            & Energy      & Source    & Spectral & Temporal \\
        & & & calibration & detection & analysis & analysis \\
          \hline\noalign{\smallskip}
        \texttt{XBANDCP-MXT}  & X-band & CP & X & X & X & X\\
        \texttt{XBANDGP-MXT}  & X-band & GP, ToO-Nom & X & X & X & \\
        \texttt{XBANDTOOMM-MXT}  & X-band & ToO-MM & X & X & X & \\
        \texttt{XBANDTOOEX-MXT}  & X-band & ToO-EX & X & X & X & \\
        \texttt{XBANDCAL-MXT}  & X-band & Calibration & X &  &  & \\
\noalign{\smallskip}\hline\noalign{\smallskip}
        \texttt{XVHFCP-MXT} & X-VHF & CP & & X & &\\
        \texttt{XVHFTOOMM-MXT} & X-VHF & ToO-MM & & X & &\\
        \texttt{XVHFTOOEX-MXT} & X-VHF & ToO-EX & & X & &\\
        \noalign{\smallskip}\hline
        \noalign{\bigskip}        
    \end{tabular}
\end{table*}

%%%%%%%%%%%%%%%%%%%%%%%%%%%%%%%%%%%%%%%%%%%%%%%%%%%%%%%%%%
\section{General workflow}
\label{sect:workflow}

The MXT pipeline has been developed using Python, in line with the approach adopted for a significant proportion of the FSC software suite. It also incorporates elements of the legacy HEASoft software\footnote{\url{https://heasarc.gsfc.nasa.gov/docs/software/lheasoft/}} \citep[][]{2014ascl.soft08004N}.
The FSC has developed a generic task scheduler, \texttt{Pipeline-Bricks}, which is capable of executing any workflow that is described in a JSON file with a schema outlined below. The workflow execution is controlled by a REST API, which is standard across all FSC pipelines. This API governs various functions, including start, kill, log access, status access, and more.

Each pipeline instance is operated as a Docker image deployed in the FSC infrastructure. The processes are triggered by a global orchestrator (see Louvin et al., RAA, 2025, 25, this issue) each time a complete set of requested input files is made available in Science Database (SDB).
For the MXT pipelines, the triggering event is the ingestion in SDB of a L1 raw event list (\texttt{MXT-EVT-CNV} for X-band, or L1 photon list (\texttt{OBPHOTL\_MXT} or \texttt{OBPHOTMM\_MXT}) for X-VHF).

Recent and ongoing processes of various pipeline instances can be monitored from a FSC user interface, displaying their status, results, and previews (see Sect.~\ref{sect:discussion_previews}).
The processing workflow is defined by a sequence of individual tasks, each one of which implemented as a specific Python module. Although there are similarities, the modules included and details of task sequences are specific to each programs and data streams. Table~\ref{tab:pipelines} lists the existing pipeline instances and their properties. Each workflow is configured in a designated JSON file, which outlines the sequence of tasks, the input data and parameters, and the output files.

The processing sequence comprises about 40 independent tasks. These tasks communicate with each other via a JSON file named \texttt{runtime\_data.json} containing global information (e.g. observation identifier) and other user data. This file is also used to communicate with the scheduler, for example to explain why a crash occurred or which preview should be stored in the host Docker container.

\begin{figure*}
   \centering
   \includegraphics[width=\hsize]{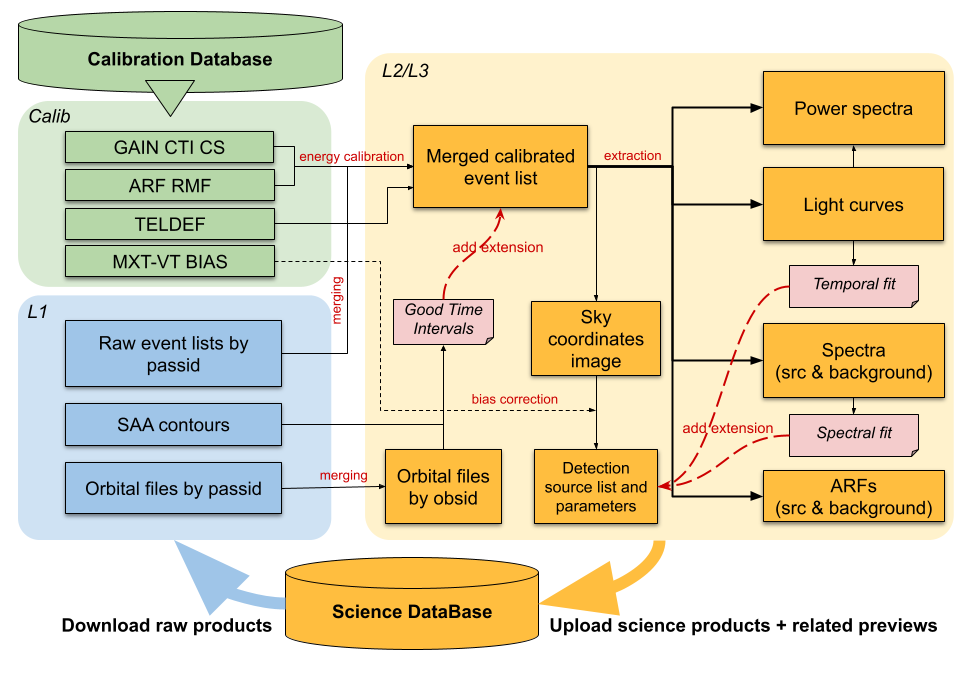}
   \vspace{-2.0em}
   \caption{Flow chart of the MXT pipeline. Coloured arrows indicates input/output to calibration and science data bases. Files and products are separated by their processing level (e.g. L1 vs L2 or L3). All the Scientific Data Products of the pipeline are represented in the orange rectangles.}
   \label{fig:pipeline_flowchart}
\end{figure*}

Figure~\ref{fig:pipeline_flowchart} presents a flowchart of the pipeline, including the input/output databases and the various stages of processing that are described in Sect.~\ref{sect:stages}.

We also provide a script that enables users to run offline analysis locally. This script builds and runs a local Docker image that incorporates the entire pipeline. Users can set several processing parameters such as custom start and stop times, binning time for light curve, target significance and maximum number of detected sources in the detection. Products generated by this local instance are only stored locally. This tool enables end users to test different setups for data analysis, for example highlighting data features that would not have been detected by regular processing.

% \todo{Add possibility of user running pipeline locally ("offline" analysis) ?}

%-------------------------------------------------------------

%%%%%%%%%%%%%%%%%%%%%%%%%%%%%%%%%%%%%%%%%%%%%%%%%%%%%%%%%%
\section{Pipeline stages}
\label{sect:stages}

\subsection{Data access and preparation}
\label{sect:stage_preparation}

In this first stage, necessary data are pulled from the relevant databases, and formatted for further processing. 
L1 event lists for a given observation, identified by its \texttt{OBS\_ID} and additionally its burst identifier \texttt{BURST\_ID} in Core Program (hereafter just "\textit{identifier"}), are split according to the X-band pass they were received from, and by the position of the MXT filter wheel. For sky observation the used filter wheel positions are OPEN and FILTER, even if no filter is physically mounted on either positions (Moita et al., RAA, 2025, 25, this issue). All the raw event lists for a given \textit{identifier} are downloaded from the SDB, concatenated, and sorted by increasing event time. The time and date of beginning and end of the merged event list is reconstructed from the set of \texttt{TSTART}/\texttt{TSTOP} keywords of individual event lists. The merged event list is renamed to \texttt{MXT-EVT-CAL} for \textit{calibrated} event list, on which all subsequent processing will be performed.

Next, all ancillary attitude and orbital files covering the observation time are downloaded from the SDB, and similarly concatenated and sorted by increasing time. If the time of an event is not matched by an attitude file, it cannot be registered on the sky and used in the analysis. Hence, the coverage fraction of events by ancillary file is calculated. Below 90\,\% coverage, the pipeline is put in delayed mode, and will restart automatically at later time, until coverage increase as more ancillary files become available in subsequent X-band passes. Above 90\,\% coverage the processing continues normally, but uncovered events are not used in the analysis. The "aspect" of the observation, akin to the mode of the satellite pointing\footnote{right ascension, declination, and roll angle.} during the MXT exposure, is calculated with HEASoft task \textit{aspect} and saved in the processing \textit{runtime} file.
Finally, the South-Atlantic-Anomaly (SAA) file, describing the times of SAA passage for SVOM/MXT in the week of the observation, is also pulled from the SDB.
For Core Program observation, the trigger time is retrieved and saved in the processing \textit{runtime} file, to calculate the delay between trigger and MXT data.

The pipeline then obtains all the calibration files describing the energy calibration, MXT optics and alignment, and spectral responses, required for the analysis, from the calibration database (CALDB). These files are maintained by the MXT Instrument Center (MIC), and the pipeline downloads the most up-to-date version of each file.

%-------------------------------------------------------------
   \begin{figure*}[t]
   \centering
   \includegraphics[width=\hsize]{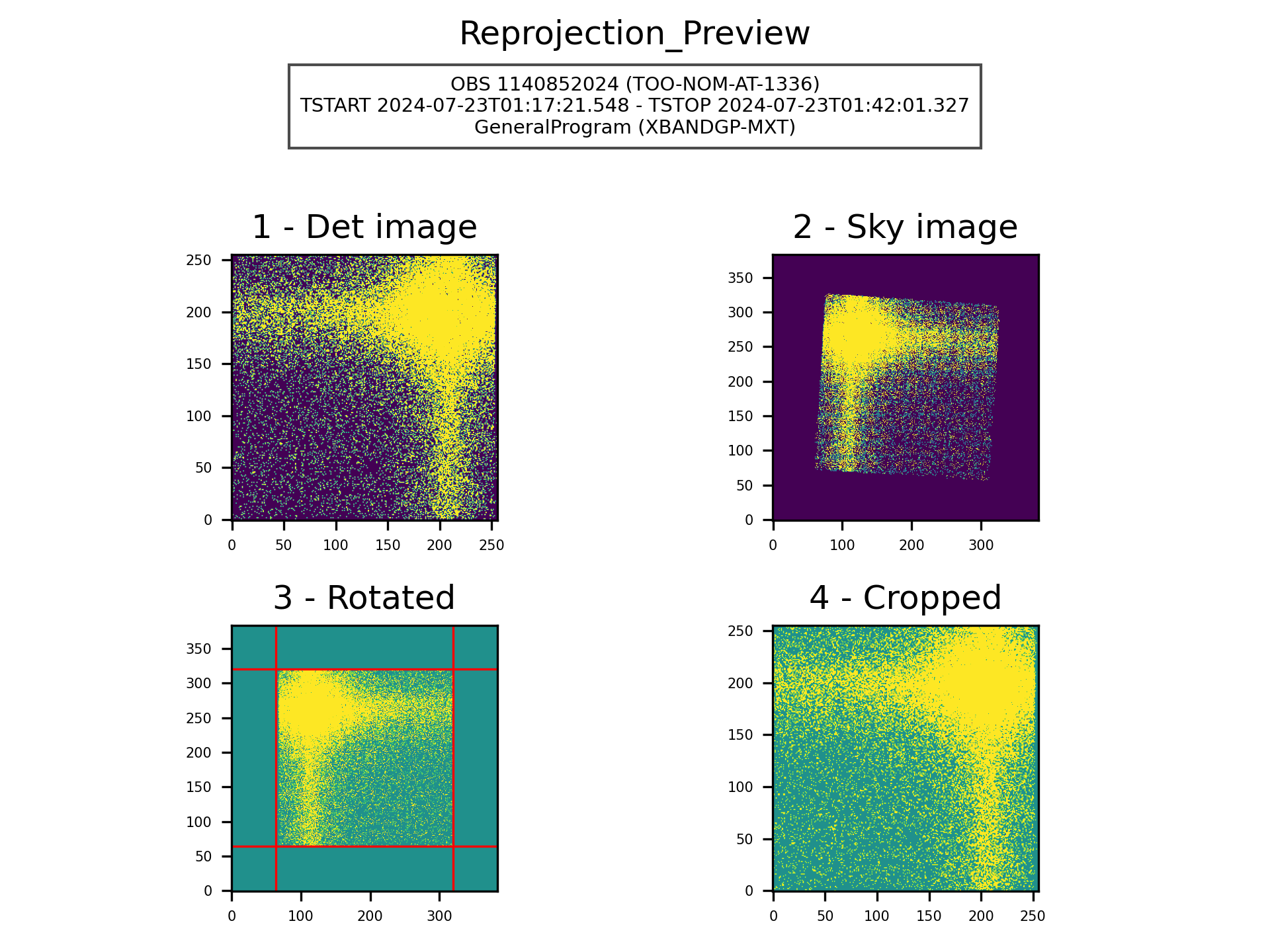}
   \caption{\textit{Left:} Illustration of de-projecting the sky image (2) to "pseudo-detector" coordinates image (4), on which source detection is performed. \textit{Right:} Histogram stage of source detection, fitting a 1--D PSF along each cross-direction.}
   % From Cyg X-1 september 2024 bias calibration, obsid 1140852024
   \label{fig:reprojected_image}
   \end{figure*}

\subsection{Events calibration and filtering}
\label{sect:stage_calibration}

\subsubsection{Energy calibration}
\label{sect:stage_calibration_energy}
The energy of each pixel hit in an event is reconstructed from the input \textit{phas} values, applying the gain and offset correction measured by the MIC for each CCD column (\texttt{RAWY}), to obtain values in keV. The energies are then corrected for the additional row-dependent effect introduced by the updated reset anode voltage\footnote{See Moita et al. (RAA, 2025, 25, this issue) for a description of this and other camera effects.} (with a factor depending on the CCD row (\texttt{RAWZ}), and the charge transfer inefficiency (CTI) depending on the pixel hit energy and number of CCD transfers to the readout (and hence on \texttt{RAWZ}). Next, these individual energies in the event pattern are summed, and corrected for the charge-splitting effect, due to the charge cloud of an X-ray hit extending to neighbouring pixels. If the charge in neighbouring pixel is above the readout threshold it can be recorded as part of the same event with the appropriate pattern (\texttt{GRADE}). If this charge is however less than the readout threshold, it is not recorded and lost. Thus, the event energy is multiplied by an energy-dependent factor ($> 1$) that depends on its \texttt{MULTIPLICITY}. Quadruple events are considered to encompass the whole charge cloud for a genuine X-ray interaction and are not affected. Events with higher multiplicity are considered invalid (generally coming from non-X-ray interactions).

Finally the summed and corrected energy of each event is discretised in the energy channels of the redistribution matrix file (RMF), and stored in the column \textit{PI} (for "pulse-invariant"), following the convention of OGIP CAL/GEN/92-002. Events with energy outside the calibrated energy range of 0.1~keV to 10~keV are assigned a null value of \texttt{PI} = -1 and are/shall not be used for analysis. For completeness and ease of use, the energy in keV is also stored in the column \textit{ENERGY}.

\subsubsection{Sky registration and image}
\label{sect:stage_calibration_registration}
The position of each event on the detector is recorded in (\texttt{RAWY}, \texttt{RAWZ})\footnote{At system level the +X direction represent the spacecraft pointing direction} form at a given time \texttt{TIME}.
Sky coordinates (\texttt{X}, \texttt{Y}) are calculated for each event, using the attitude file to get the satellite pointing coordinates at the same time (interpolating if need be) and taking into account the detector and telescope orientation.

\subsubsection{Time and pixel filtering}
\label{sect:stage_calibration_filtering}
Good-time-intervals (GTI, OGIP Memo OGIP/93-003) are prepared to delineate various effects:
\texttt{MXT-GTI-OBS} combines the observation times of each raw event lists (Sect.~\ref{sect:stage_preparation}); 
\texttt{MXT-GTI-ELV} to limit observations above a critical angle between pointing and Earth-limb. Due to straylight conditions, observations are only performed above an angle that varies between day and night time (Robinet et al., RAA, 2025, 25, this issue), hence this critical angle is 0 by default and thus marks the period of Earth occultation; and
\texttt{MXT-GTI-SAA} to outline the periods outside of SAA passage.
Additionally, users reprocessing data may specify a custom time period, relative to the observation start time, for selecting data (\texttt{MXT-GTI-USER}).
All these GTI files are combined with a logical "AND" and stored in the \texttt{GTI} extension of the calibrated event lists and SDPs.
In CP mode, a GTI for the analysis of the "early" period data only is also created (\texttt{GTI\_EARLY}) and saved in the same files. The default length for the "early" period is hard-coded to 1200~s in the pipeline, which will usually coincide with the first orbit of follow-up observation.
The effective exposure time summing all GTIs are calculated and stored in the files.

Bad pixels are identified on a detector-coordinates image using 3-$\sigma$ clipping over a 40$\times$40 pixels sliding window. The search is ran separately on a 0.6-10~keV energy band image (standard bad pixels), and a low energy 0.1-0.6 keV band to identify so-called "rain effect" bad pixels (Moita et al., RAA, 2025, 25, this issue). The latter are spurious and transient bright pixels that affect several (a few to a few dozen) pixels of the same column. As they are bright and confined to lower energy, they can be effectively identified and flagged from the 0.1-0.6~keV image. Events from identified bad pixels are \textit{not} removed from the calibrated event list. Instead, bad pixel locations are stored in \texttt{BADPIX} and \texttt{BADPIX\_RAIN} extensions and in the processing \textit{runtime} file. All SDPs are produced with bad pixels filtered out.

\subsection{X-ray image and source detection}
\label{sect:stage_detection}

A sky-coordinate image (\texttt{MXT-SKY-IMA}) is created from the \texttt{X} and \texttt{Y} columns of the events, using a standard energy band of 0.3~keV to 10~keV, single- to quadruple-pixel events, and filtering out bad pixels. The image is a gnomonic projection \citep[\texttt{TAN},][]{2002A&A...395.1077C} in equatorial coordinates (right ascension and declination) with reference coordinates at the observation aspect point calculated earlier (Sect.~\ref{sect:stage_preparation}). 
Due to the properties of the lobster-eye optics, the PSF is aligned with the \textit{detector} axes, and a detector-coordinates image should be the natural basis for source detection. However, the satellite pointing may change during the observation: For instance during Core Program follow-up of a GRB, if the onboard software identifies a candidate afterglow, the satellite will slew a second time to put this source at the centre of the MXT and VT field-of-view. The source would appear at two positions on the time-integrated detector-coordinate image.
Thus, the pipeline "de-projects" the sky image (with all sources at the same \texttt{X}/\texttt{Y} position), correcting the satellite roll angle and symmetries back to "pseudo-detector" coordinates, and cutting the $256\times256$ pixel image around the centre. This process is illustrated in Fig.~\ref{fig:reprojected_image}.

%-------------------------------------------------------------
\begin{figure}
   \centering
   \includegraphics[width=\hsize]{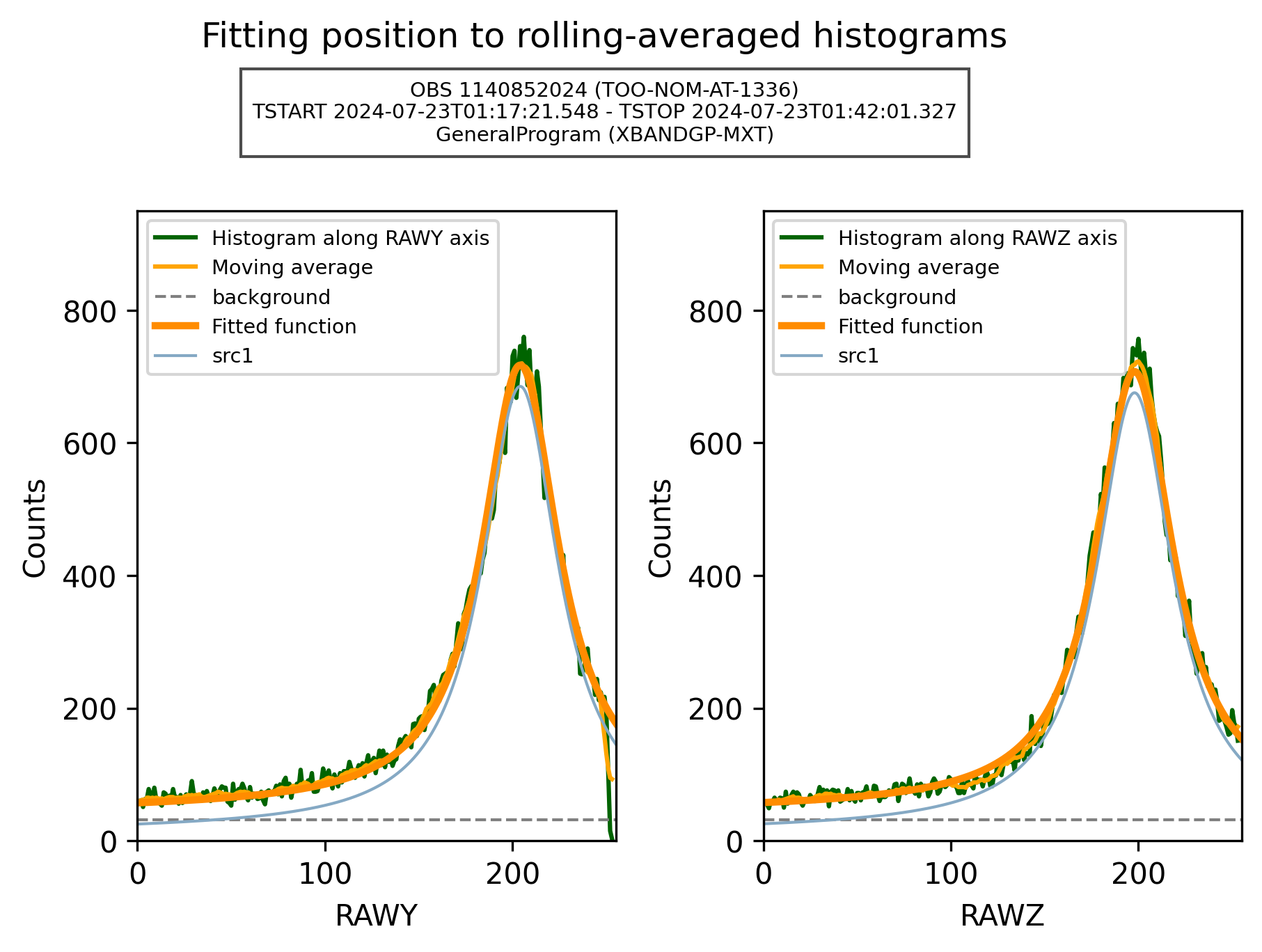}
   \caption{Histogram stage of source detection, fitting a 1--D PSF along each cross-direction.}
   % From Cyg X-1 september 2024 bias calibration, obsid 1140852024
   \label{fig:histogram_detection}
\end{figure}

Source detection is performed on this image using a maximum-likelihood 2D PSF-fitting algorithm. Iteratively, the significance of finding $N$ source(s) in the image is computed from the change in best-fit $C-$statistics \citep{1979ApJ...228..939C} compared to $N - 1$ source(s). For $N=1$, it is compared to the $C-$statistics of a flat image with a fitted uniform background count rate, requiring a significance corresponding to a 99\,\% confidence level. (C.L.).
The assumption of a flat background is valid for blank sky in normal conditions, but breaks down in cases of adverse space weather. In such cases spurious sources may be detected and/or sensitivity degraded. Identification and treatment of these cases is reserved for future pipeline updates.
The pipeline iterates until new sources are no longer significant at a higher C.L., chosen to avoid the presence of spurious sources in the cross-arms of the first source(s).

The initial position and normalisation of the $N$ sources are identified by fitting a one-dimensional integral of the PSF (Feldman et al., RAA, 2025, 25, this issue) to a rolling-average of the histogram of the count rate along each cross direction (Fig.~\ref{fig:histogram_detection}). In a second stage, the initial position and normalisation are optimized by fitting a PSF model of the $N$ source(s) to the input image, maximising the $C-$statistics. Although a single image and thus energy band is used, the model PSF is energy-weighted, using the ratios of count rates in each of the 11 energy bands at which the PSF was calibrated to the total count rate.

Uncertainties on the position (in each direction) and in normalisation are obtained by varying them and finding the values where change in  $C-$statistics corresponds to a 90\,\% C.L. for one parameter of interest (i.e. $\Delta C = 2.706$).
The source position is corrected for alignment bias using the MXT-VT bias matrix, prepared by the MIC from observations of bright sources at nine positions covering the detector. This procedure is described in detail in G\"otz et al. (RAA, 2025, 25, this issue).
The positions in detector and sky coordinates, normalisation, and uncertainties are recorded in the SDP called \texttt{MXT-SOP-IMA} (for SOurce Parameters).

\subsection{Source products}
\label{sect:stage_sourceproducts}
Light curve, energy spectrum, and power spectrum (\texttt{MXT-SRC-LC}, \texttt{MXT-SRC-SP}, and \texttt{MXT-SRC-PS}, respectively) are extracted for each detected source, using only valid X-ray grades, excluding bad pixels, and in the calibrated energy band of 0.3~keV to 10~keV. The source extraction regions are the union of a circle of 50 pixel radius, and two rectangles of 50 pixel width along the principal axes of the detector, all centred on the detected position. These encompass the central patch and cross-arms of the PSF, respectively. Background extraction regions are the remaining part of the detector \textit{not} covered by a source region.
A source-specific effective ancillary response file (\texttt{MXT-SRC-ARF}) is created for each detection to store the effective area, taking into account the number of "rain" bad pixels within the source extraction region. The effective area below 0.6~keV is thus reduced by the corresponding number of excluded pixels.

Energy spectra are not rebinned and retain the native binning of the redistribution matrix file which uses 1024 uniform bins between 0.1 and 10~keV (i.e. bin widths are just below 10~eV). Count rate in light curves has default time bins of 10~s.
These SDP are provided for all X-band programmes except for calibration observation (with closed filter wheel). Automatic analyses are performed as listed in Table~\ref{tab:pipelines} and described in the following.

\subsubsection{Spectral analysis}
\label{sect:stage_sourceproducts_spectral}
Energy spectra are analysed with XSPEC \citep{1996ASPC..101...17A} using single-component models: Power law (\texttt{pow}), black body (\texttt{bbody}), and collisional-ionisation equilibrium thermal model (\texttt{apec}), absorbed using the Tuebingen-Boulder ISM absorption model \texttt{TBabs}. Fit is performed over the 0.3~keV to 10~keV band on non-rebinned data using the Cash fit statistics \citep[C-stat,][]{1979ApJ...228..939C}. Best-fit parameters and associated 90\,\% C.L. uncertainties are saved in the source detection and parameters file \texttt{MXT-SOP-IMA}. If either the fit or uncertainty calculation fails for a given model, values are set to null.

%-------------------------------------------------------------
   \begin{figure*}
   \centering
%   \hspace{-4.5em}
   \includegraphics[width=\hsize]{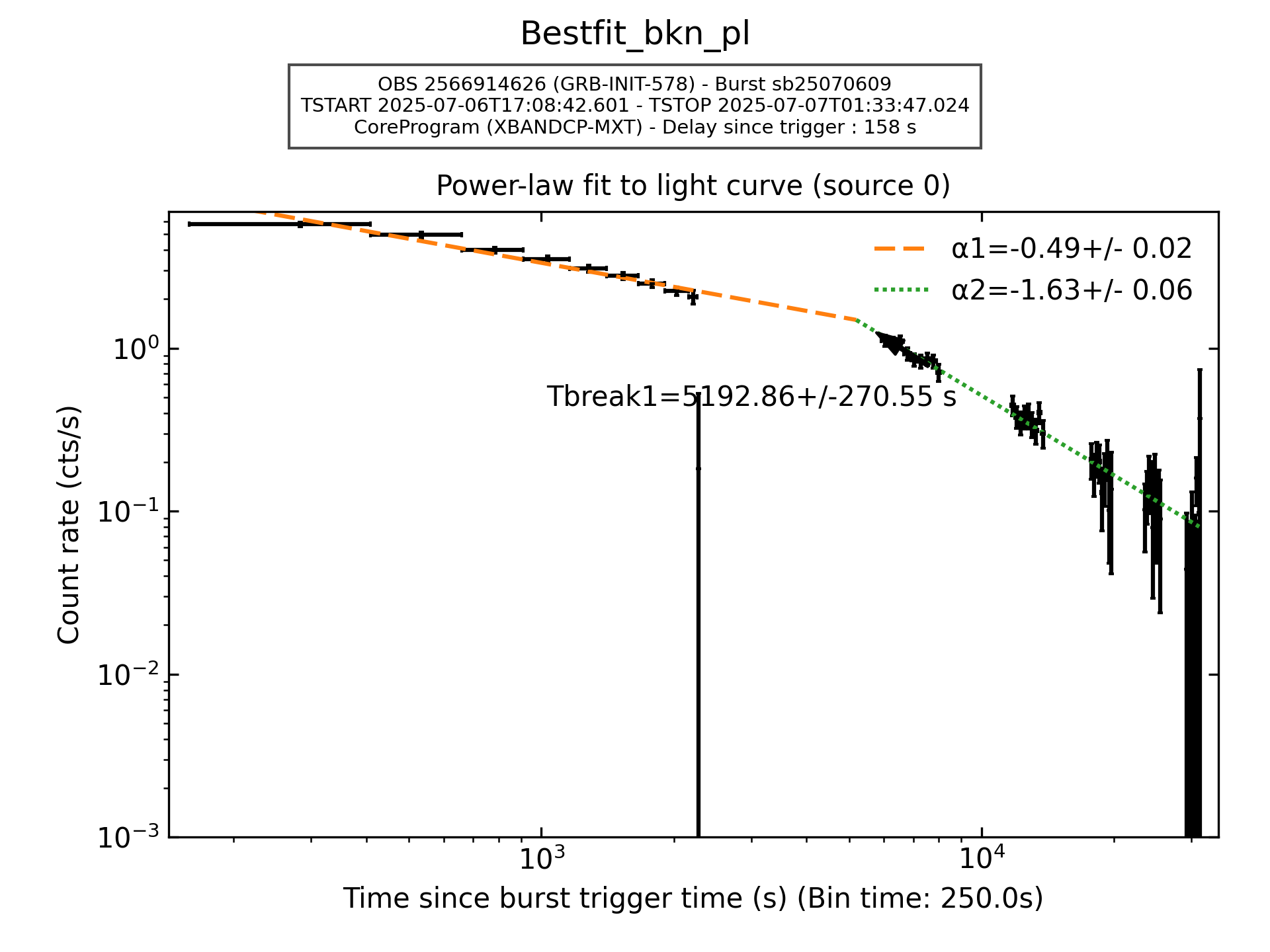}
   \caption{Preview of the light curve from a CP observation of GRB250706B/C with segmented power-law fit.}
   % From GRB250706 with 250s bin, burstid sb25070609
   \label{fig:products_LC}
   \end{figure*}

%-------------------------------------------------------------
   \begin{figure*}
   \centering
   \includegraphics[width=\hsize]{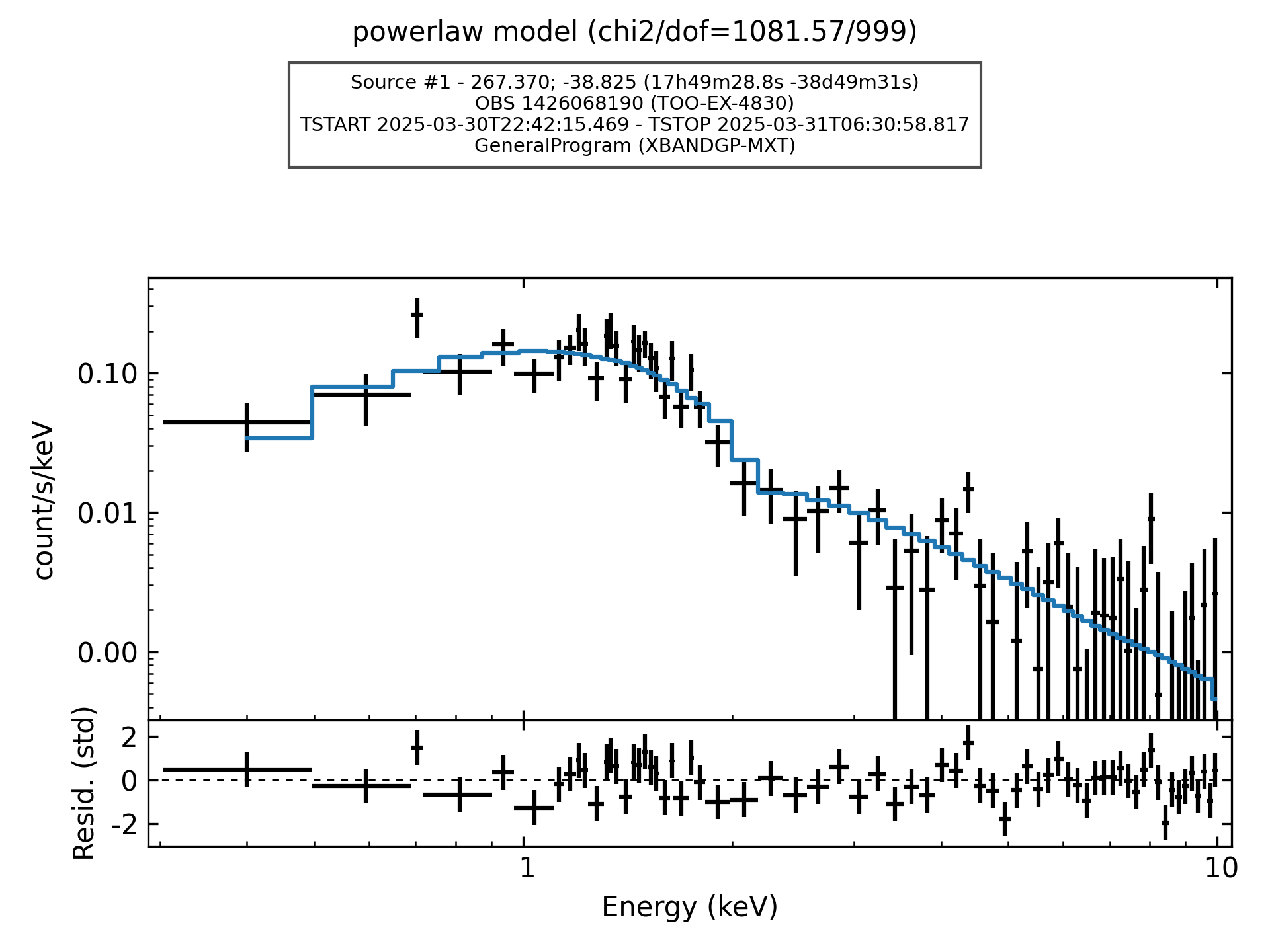}
   \caption{Preview of a spectrum from a ToO-EX observation with power-law model fit.}
   % From obsid 1426068190
   \label{fig:products_SP}
   \end{figure*}

\subsubsection{Temporal analysis}
\label{sect:stage_sourceproducts_temporal}
While binned light curve are produced for all detected sources in each science modes (Table~\ref{tab:pipelines}), only Core Program sources have an automatic temporal analysis. This analysis is geared towards GRB afterglows, which in general show fading light curves. We use segmented power-laws to model the afterglow light curve \citep{2009MNRAS.397.1177E}. The number of possible segments is restricted to three in our case, as was found appropriate given the count rate expected with MXT. No automatic detection of potential flare is attempted. Fitting is performed using non-linear least-squares with bound parameters and a Trust Region Reflective \citep[TRF,][]{Coleman1994-vo} algorithm for minimisation. The minimal number of bins is verified before attempting to use $N>1$ segments in the light curve.

The light curve parameters are given relative to the reference time, which is the trigger time by default. If said trigger time cannot be found, the reference time is the MXT data \texttt{TSTART}.
Best-fit temporal indices, break times between segments, and associated uncertainties, are added to the source detection and parameters file \texttt{MXT-SOP-IMA}. An example of automatic GRB light curve analysis is shown in Fig.~\ref{fig:products_LC}

\subsection{Upload of products}
\label{sect:stage_upload}

In the final stage of the pipeline, the products are uploaded to the SDB. Observation-level products (calibrated event list, sky-coordinates image, source list) are uploaded for each observation. Source-specific products (Sect.~\ref{sect:stage_sourceproducts}) are only uploaded if the detection list is non-empty. For X-VHF data the only products are the sky-coordinates image and source list, due to the limited photon information (Sect.~\ref{sect:datastreams_xvhf}).
If a new processing uploads a SDP with a different pipeline version or data model, this is considered a new product with a new product identifier and an incremented product version. Otherwise it is considered an update of the same existing product, and only the update number field is incremented.

%%%%%%%%%%%%%%%%%%%%%%%%%%%%%%%%%%%%%%%%%%%%%%%%%%%%%%%%%%
\section{Discussion}
\label{sect:discussion}

\subsection{Product previews}
\label{sect:discussion_previews}
Science Data Products are in the FITS format. For easier visualization of the products and pipeline output, the pipeline also generates previews, most often in standard \textit{png} image format. These include images in sky coordinates, energy spectra with and without spectral models fitted, various visualizations of the light curve with and without fits, etc. Previews of tabular data such as source list parameters are created in JSON format.

Previews from recent pipeline processes are visible in the pipeline monitoring user interface hosted at the FSC. They are also stored in the SDB : Each preview is associated to a unique SDP, but one SDP may have multiple previews attached to it. All associated previews of an SDP are available on the SDB as "ancillary files", which are grouped together in a single PDF file. Previews display metadata allowing to trace them back to their origin: Observation or burst identifier(s), SVOM program and pipeline instance, time interval of the data used, source identifier and celestial position.
Examples of previews associated to source detection (\texttt{MXT-SOP-IMA}) are shown in Fig.~\ref{fig:reprojected_image}, while previews for light curve and energy spectrum (\texttt{MXT-SRC-LC} and \texttt{MXT-SRC-SP}, respectively) are in Figs.~\ref{fig:products_LC} and \ref{fig:products_SP}.

\subsection{Virtual Observatory compliance}
\label{sect:discussion_VO}

Publishing scientific level products in the Virtual Observatory (VO) is an efficient way to widen their dissemination: \textit{i)} this allows them to be dynamically merged, cross-matched or overlaid with datasets taken out from a very huge number of resources and \textit{ii)} data published in the VO are declared in a public registry so that they can be retrieved by using any sort of filter (mission, energy band, sky coverage\ldots). Being present in the VO makes SVOM data available for any astronomer in any scientific context.
To this end, all L2/L3 SDPs have a specific extension named \texttt{VO-TAGS} that contains keywords mapping directly to the fields of the ObsCore \citep{2017ivoa.spec.0509L} data model, facilitating publication of the product in the Virtual Observatory, particularly in the ObsCore table of TAP services \citep{2019ivoa.spec.0927D}.

\subsection{Provenance and reprocessing}
\label{sect:discussion_provenance}
The aforementioned \texttt{VO-TAGS} extension also contains a description of the sequence of tasks (with their parameters) used to build the product in question. This description can be interpreted by a specific instance of the pipeline in reprocessing mode, which can then rebuild a product after the scientist user has tuned some parameters.

\section{Conclusion and perspective}
\label{sect:conclusion}

The MXT data analysis pipeline described herein has been running automatically at the FSC since the launch and commissioning phase of SVOM.
Identified bugs have been corrected and new features added since, accounting for unexpected effects such as the MXT "rain effect". Products and analyses are still being improved, taking into account feedback and feature requests from users and scientists. New features and bug fixes are tracked in change logs on the  GitLab instance hosting the pipeline code (Louvin et al., RAA, 2025, 25, this issue).

Products from this pipeline are the basis for all analyses involving MXT:
\textit{i)} X-VHF data are independent confirmation of the onboard software analysis and is valuable for the rapid localization of GRB afterglows (e.g. General Coordinates Network Circulars with MXT position);
\textit{ii)} X-band data products enable the detailed study of early GRB afterglow light curves (e.g. GRB 250317B, Zhao et al. in prep.), or allow scientists to estimate X-ray upper limits shortly after the trigger for high-redshift bursts \citep{2025arXiv250718783C}. In General Program MXT pipeline products allow the study of non-GRB X-ray transient both Galactic (e.g. doublet burst in EXO 0748$-$676, Rawat et al., in prep.) and extragalactic \citep[][Foisseau et al., in prep.]{2024GCN.38452....1G,2024ATel16935....1C}. Finally, it also contributed to multi-messenger astronomy, analysing the MXT tiles covering the footprint of gravitational wave events or ultra-high energy particles, for instance in the case of the 220~PeV neutrino \citep[][ and follow-up thereof, \citealt{2025arXiv250208484K}]{2025Natur.638..376T}.

The MXT data analysis pipeline performed satisfactorily on data from all data streams and programs of the mission, conducting the necessary calibration, processing, and analysis, and provides science-ready products to the scientists. Thus, the pipeline is an integral part of the successful run of the SVOM mission in all its intended programs.

\begin{acknowledgements}
The Space-based multi-band astronomical Variable Objects Monitor (SVOM) is a joint Chinese-French mission led by the Chinese National Space Administration (CNSA), the French Space Agency (CNES), and the Chinese Academy of Sciences (CAS). We gratefully acknowledge the unwavering support of NSSC, IAMCAS, XIOPM, NAOC, IHEP, CNES, CEA, and CNRS.
The authors acknowledge the strong involvement of the CNES team during the commissioning phase  (K. Mercier, A. Fort, S. Crepaldi).
\end{acknowledgements}

% \appendix                  %%appendicial material is supported

\bibliography{msRAA-2025-0575_R1-bibtex}

\label{lastpage}

\end{document}